# THE AHARONOV-BOHM EFFECT IN THE FRACTIONAL QUANTUM HALL REGIME

J.D.F. Franklin[†], I. Zailer, C.J.B. Ford, P.J. Simpson, J.E.F. Frost, D.A. Ritchie, M.Y. Simmons and M. Pepper

*Cavendish Laboratory, University of Cambridge, Madingley Road, Cambridge, CB3 0HE, UK*

We have investigated experimentally resonant tunnelling through single-particle states formed around an antidot by a magnetic field, in the fractional quantum Hall regime. For 1/3 filling factor around the antidot, Aharonov-Bohm oscillations are observed with the same magnetic field period as in the integer quantum Hall regime. All our measurements are consistent with quasiparticles of fractional charge $e^*$. However, the results are also consistent with particles of *any* charge ($\geq e^*$) as the system must rearrange every time the flux enclosed increases by $h/e$.

The Fractional Quantum Hall Effect (FQHE) is a much-studied many-body effect [1]. Complex electron-electron interactions give rise to a quantized Hall resistance, and to oscillations in the longitudinal resistance of a high-mobility two-dimensional electron gas (2DEG). For particular fractional values $p/(2p+1)$ of the filling factor $\nu$ (the ratio of the number of conduction electrons to magnetic flux quanta $h/e$), where $p$ is an integer, we can change the basis set of the states considered to quasiparticles having fractional charge $e^* = e/(2p+1)$ [2-4]. For example $p = 1$ implies quasiparticles of charge $e^* = e/3$.

Observation of fractional Hall plateaux is not, however, direct evidence for the existence of such quasiparticles [5-7]. It was suggested [5] that the Aharonov-Bohm (AB) effect might be used to determine $e^*$, since the single particle (SP) states around an antidot might be expected to enclose multiples of a flux quantum $h/e^*$ instead of $h/e$. At first this appeared to be experimentally supported [8], but then it was pointed out that gauge invariance enforces a maximum period of $h/e$ [6,9]. Thus current theories predict an $h/e$ period in resonant tunnelling through an SP state in the FQHE regime for fractionally charged quasiparticles.

Further, Coulomb blockade (CB) of fractionally charged quasiparticles localized around the antidot might be expected to have an energy scale $e^{*2}/C_d$, $C_d$ being the capacitance of the antidot states to everything else [9]. Alternatively the SP energy spacing should scale as $E_r e^*/B$ where $E_r$ is the slope of the potential at the SP state. The $h/e^*$ periodicity might obtain only if gauge invariance could be broken by shifting the system sufficiently adiabatically on a timescale apparently unobtain-ably short [6].

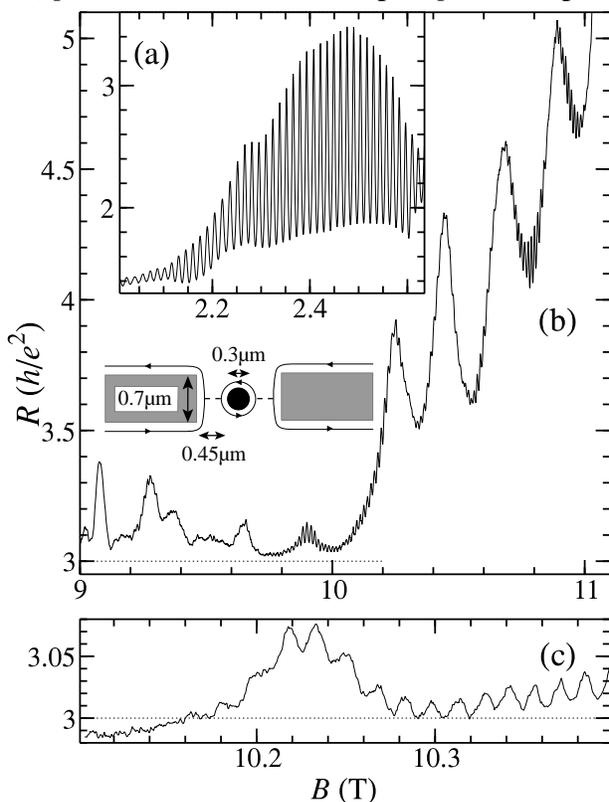

**Figure 1.** Aharonov-Bohm oscillations in the integer (a) and fractional (b) QHE regimes; (c) shows an enlargement of the $3h/e^2$ plateau region similar to that in (b). Dotted lines indicate the plateau (Sample A).

[†]E-mail: jdff1@cam.ac.uk

In this paper we present measurements of the period of AB oscillations in the integer and fractional regimes. We show that the period in magnetic field is the same in both regimes for four different devices. This period is consistent with being $h/e$ from the defined size of the antidots. We have also studied the period in gate voltage in the two regimes [10], and deduced single particle energy-level spacings from temperature dependences and DC bias measurements. Our device consists of a patterned 2DEG formed in a GaAs/Al$_x$Ga$_{1-x}$As heterostructure. An antidot (potential hump) is formed by a central 0.3 µm diameter gate, contacted separately using a special technique [11,12]. The macroscopic edge states are brought close enough to tunnel to the central states at two constrictions, the widths of which are determined by side-gate biases and magnetic depopulation (inset Fig. 1). Thus resonant tunnelling can take place through the ladder of SP states, each enclosing an integer number of $h/e$ magnetic flux quanta, allowing backscattering from one edge to the other and giving rise to oscillations in the resistance (Fig. 1) [11].

The carrier concentration $n_s$ after illumination of the wafers used was $1–2 \times 10^{15}$ m$^{-2}$ with mobilities 150–300 m$^2$V$^{-1}$s$^{-1}$. The depth to the 2DEG was 100 nm (devices A and B), 300 nm (device C) and 170 nm (device D). Four-terminal measurements were made using standard ac techniques in a dilution refrigerator, at a base electron temperature of about 30mK (indicated by a significant temperature dependence down to base). A constant current of 1nA (0.03nA) in the integer (fractional) regime was used, low enough to avoid electron heating.

We have compared the AB oscillations in the FQHE and IQHE regimes (Fig. 1). In the fractional regime, the filling factor in the bulk, $\nu_b = 2/3$, and that in the constrictions, $\nu_c = 1/3$. We can calculate the filling factor in the wider constriction in the IQHE from $\nu_c = h/e^2R$, where $R$ is the two-terminal longitudinal resistance [13], and the result is also found to hold in the FQHE. The oscillations can be seen to be rising off, and associated with, a well-defined $\nu_c = 1/3$ plateau indicating resonant reflection through a $\nu = 1/3$ state encircling the antidot. We can also, in some cases, see well-defined plateaux at other values, corresponding to $\nu_c = 2/5$ and higher while keeping $\nu_b = 2/3$ (Fig. 2), showing reflection of fractional edge states [14]. We have also observed AB oscillations for $\nu_c > 1/3$, due either to resonant transmission through inter-edge state scattering, or to transmission through intra-edge state scattering. Both processes, however, probe SP resonant tunnelling via a magnetically bound state at the Fermi energy.

We use our independent control of the central dot to vary the antidot size. AB oscillations are also seen with gate voltage $V_g$, in both integer and fractional regimes (Fig. 3). Thus we see the period $\Delta V_g \propto \nu_b$. (Devices A and B showed a deviation at high $B$ [10], which we believe was an artefact due to Coulomb blockade in the dot metallisation itself.) Devices C and D avoided this problem by a single-stage dot metallisation.

If $\nu_c$ is an integer, there is one electron per state. Each state encloses one more $h/e$ flux quantum than its predecessor. From counting electrons, we get $\Delta V_g = Ne/C_g$ where $C_g$ is the capacitance of one state to the dot gate, and $N$ the number of Landau levels affected by the dot. From counting states, and using $B\Delta A = h/e$, we get $\Delta V_g = n_s\Delta Ae/C_g = n_s h/BC_g$, which is equivalent in the integer regime. Thus $\Delta V_g \propto \nu_b$, and by modelling the electrostatics, even without self-consistency, one arrives at a figure within 15% of the experimental value. More accuracy is not expected due to uncertainties in the lithography.

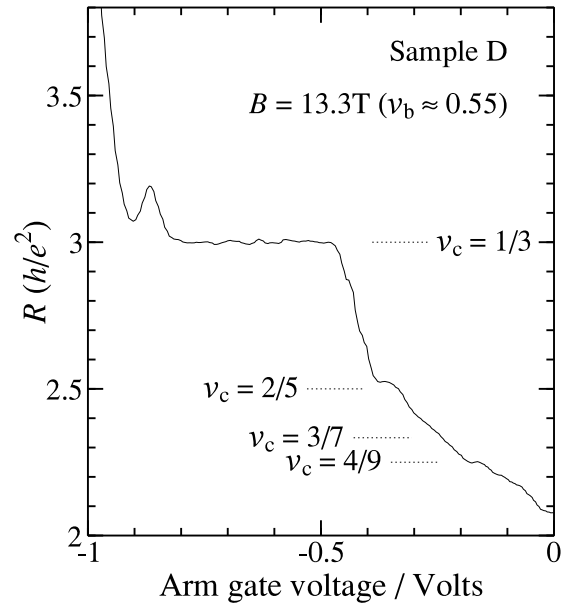

**Figure 2.** Plateaux due to reflection of fractional edge states, for one constriction.

Counting states is no longer equivalent to counting electrons in the fractional regime, however. One might expect CB of an integer number of electrons around the dot, leading to a difference between state-counting and electron-counting arguments. The states are, however, open to the Fermi sea, so normal CB arguments do not apply.

To show this, we present a qualitative interacting-electron picture (Fig. 4). Let us populate one in three of the available SP states ($\nu_c = 1/3$) with electrons. Self-consistent interaction processes force them to have a complicated many-body wavefunction, which we represent as a particular pattern (A) of occupation relative to the applied potential at the edge of the antidot. As $B$ is increased, the states move inwards to keep the enclosed flux the same.

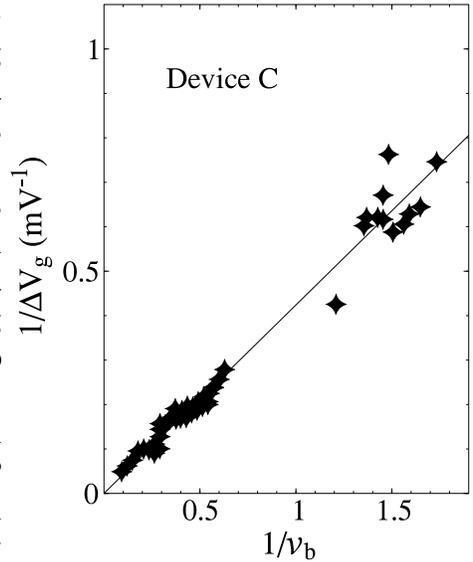

**Figure 3.** $1/\Delta V_g$ versus $1/\nu_b$ for device C. The straight line is a fit through zero to the data.

After adding an extra $h/e$ of flux, the states have shifted along by one. Suppose the electrons move with the states (picture C). This will be energetically unfavourable, because configuration A was the ground state. Instead, the system must go back to pattern A, but with each electron enclosing one more flux quantum than before. It may either relax on a very short time-scale [6], or the many-body wavefunction may itself have $h/e$ periodicity. This is the principle of gauge invariance, and it means that one counts states, not particles.

The same must happen if the potential is changed with a gate. A slight charge imbalance will appear around the antidot, but as soon as it has increased by the average charge per state, $e^*$, the system will relax to reduce this back to the original value. Thus the system will appear to consist of particles of charge $e^*$, *even if* the actual charge is greater than this, such as $0.47e$ or $e$. For composite fermions, each electron is effectively paired with two flux quanta, but at $\nu = 1/3$, there are three ways of doing this. The relaxation described is between these different pairings.

After completion of this work Goldman and Su published similar experimental results [15] that complement ours, based on calculating $e^*$ from the known capacitance to a back-gate. We have now confirmed their results. This method allows one to add single electrons or quasiparticles to the region where $\nu = \nu_c$. However, as soon as the charge has changed by $e^*$, the system relaxes again as discussed above. Thus even this technique, which appears to measure the quasiparticle charge $e^*$, only measures the average charge per state in the region to which particles are added.

The SP energy-level spacing $\Delta E$ should vary as $\nu_b \propto 1/B$. Temperature dependence and DC bias results agree that $\Delta E$ is reduced by a factor of ~3 between integer and fractional regimes, between which $B$ changed by a factor of 3.1. DC biasing in the integer (fractional) regime gave $\Delta E = 148$ μeV (46 μeV). The temperature at which the amplitude of the oscillations fell to about half its base value was $250\pm10$ mK ($60\pm10$ mK); the discrepancy may be due to the difficulty in calibrating the low electron temperatures. There is a good fit of the $T$ dependence of the area under the $h/e$ Fourier peak to thermally-broadened Fermi liquid theory (Fig. 5), and the fitted values of $\Delta E$ vary with $\nu_b$ as expected. Note that the energy scales for electrons and quasiparticles should scale with the charge. The energy of a quasiparticle under DC bias $V_{dc}$ is $e^*V_{dc}$ [16,17] and the thermal energy $k_B T$ of an electron has to be shared among the $e/e^*$ quasiparticles, giving an energy per quasiparticle of $k_B T e^*/e$.

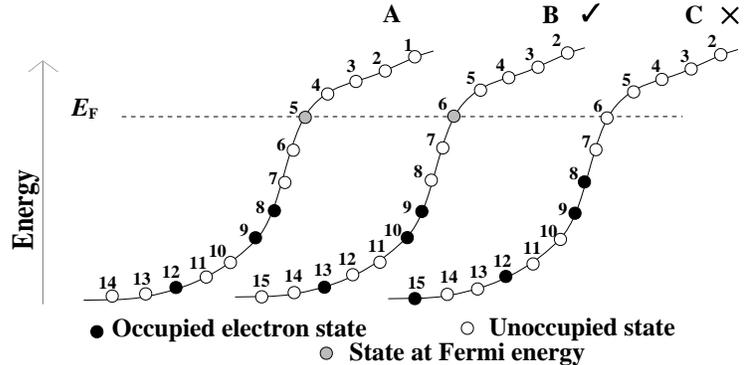

**Figure 4.** Schematic depiction of state occupation changing self-consistently as we change the potential or magnetic field. If A is a stable configuration then so is B, but not C.

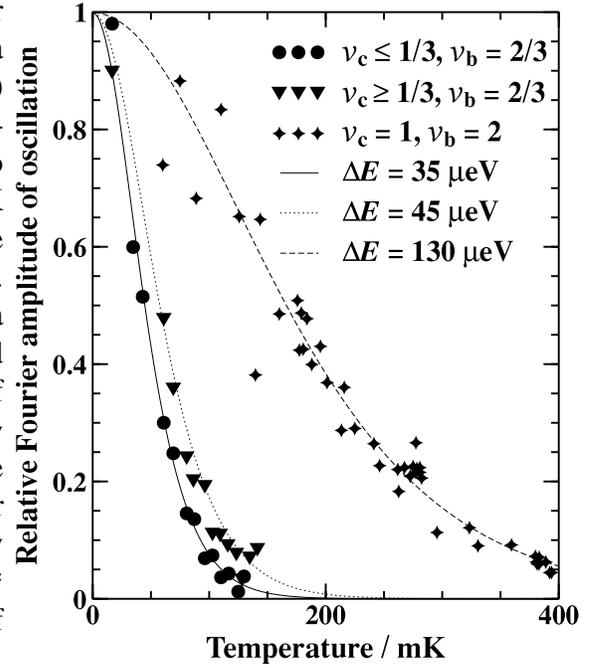

**Figure 5.** Temperature dependence of the amplitude of oscillations, and fits to thermally broadened Fermi liquid theory.

We find that the dependences in DC bias, temperature and dot voltage agree between two processes: resonant reflection via the central state, and inter-edge state resonant transmission. The two processes can be differentiated as they lead to peaks in resistance and conductance respectively. The, perhaps surprising, conclusion is that quasiparticles can tunnel through a barrier in the latter case, perhaps because the quasiparticle state is no longer completely localised, and hence the charge on it is no longer absolutely quantized in units of $e$.

We thank M. Büttiker, B.I. Halperin, G. Kirczenow, D.R. Mace, C.H.W. Barnes and M. Yosefin for helpful discussions, and acknowledge support from the UK EPSRC and Trinity College (IZ).